\documentclass[prb,showpacs,preprint,preprintnumbers]{revtex4}%
\usepackage{graphicx}
\usepackage{dcolumn}
\usepackage{bm}
\usepackage{amsmath}
\usepackage{amsfonts}
\usepackage{amssymb}%
\begin{document}
\title{Thermodynamic properties of the periodic Anderson model : X-boson treatment}
\author{R. Franco}
\email{rfranco@if.uff.br}
\author{M. S. Figueira}
\email{figueira@if.uff.br}
\affiliation{Instituto de F\'{\i}sica, Universidade Federal Fluminense (UFF), Av.
Litor\^{a}nea s/n, 24210-340 Niter\'{o}i, Rio de Janeiro, Brazil, Caixa Postal 100.093}
\author{M. E. Foglio}
\email{foglio@ifi.unicamp.br}
\affiliation{Instituto de F\'{\i}sica \textquotedblleft Gleb Wataghin\textquotedblright,
Universidade Estadual de Campinas, 13083-970 Campinas, S\~{a}o Paulo, Brazil,
Caixa Postal 6165}
\date{\today }

\begin{abstract}
We study the specific heat dependence of the periodic Anderson model (PAM) in
the limit of $U=\infty$ employing the X-boson treatment in two different
regimes of the PAM: the heavy fermion Kondo (HF-K) and the local magnetic
moment regime (HF-LMM). We obtain a multiple peak structure for the specific
heat in agreement with the experimental results as well as the increase of the
electronic effective mass at low temperatures associated with the HF-K
regime.The entropy per site at low T tends to zero in the HF-K regime,
corresponding to a singlet ground state, and it tends to $k_{B}ln2$ in the
HF-LMM, corresponding to a doublet ground state at each site. The linear
coefficient $\gamma(T)=C_{v}(T)/T$ of the specific heat qualitatively agrees
with the experimental results obtained for different materials in the two
regimes considered here.

\end{abstract}
\maketitle

\section{Introduction}

\label{Sec1}

Heavy fermions metals (HF) are intermetallic compounds of certain lanthanides
and actinides. They contain a lattice of $f$ ions, embedded in a Fermi sea of
conduction electrons $c$. These materials exhibit very large specific heats
and spin susceptibilities at low temperatures, and these characteristics are
associated with high effective masses of quasi-particles, caused by the local
strong Coulomb interaction between the $f$ electrons. The periodic Anderson
model (PAM) is considered a fairly good description of these systems, and the
relevant parameters of this model are: the hybridization $V$ between the
localized $f$-electrons and the itinerant $c$ electrons (associated with the
$s$, $p$ and $d$ orbitals), the value of the energy $E_{f}$ of the localized
$f$ \ level, the value $U$ that determines the Coulomb repulsion energy
between the $f$ electrons, the dispersion relation of the $c$-electrons and
the chemical potential $\mu$. The physics of these materials is dominated by
the competition between the Kondo effect, that tends to quench\ the magnetic
moments of the $f$ levels,\cite{Kondo} and the RKKY
(Rudermann-Kittel-Kasuya-Yosida) interaction, that tends to introduce
antiferromagnetic order in the lattice.\cite{RKKY,Mucio} Some of these
compounds develop superconducting states at low temperatures, and their nature
is not yet completely understood.\cite{Steglich1, Steglich2}.

The thermodynamic properties of the PAM have been recently studied employing
different methods: equation of motion (EOM)\cite{Lacroix, Chinos} and two
variants of the mean field theory\cite{Portuges} that respectively describe
the weak coupling limit (Hartre-Fock) and the strong coupling one (slave boson
mean field theory (SBMFT)\cite{Coleman1,Coleman2,NewnsR}). An interesting
result in all these works\cite{Lacroix, Chinos, Portuges} is the presence of a
multiple peak structure in the temperature dependence of the specific heat. In
the works that use the EOM, the first peak was associated with spin
fluctuations and the second one with charge fluctuations\cite{Lacroix,
Chinos}. Bernhard et.al.\cite{Lacroix}, defined a lattice Kondo temperature
$T_{K}$ (we call it $T_{KL}$ in this work) as equal to the $T$ corresponding
to the minimum of the temperature derivative of the average $\left\langle
c_{i\sigma}^{\dagger}f_{i\sigma}\right\rangle $, that measures the
\textquotedblleft transference\textquotedblright\ of electrons from the
localized levels to the conduction band and vice-versa. On the other hand,
Peres et.al.\cite{Portuges}, define a correlation temperature $T^{\ast}$ (we
call it $T_{0}$ in this work) equal to the $T$ corresponding to the maximum of
the temperature derivative of the parameter $\left\langle c_{i\sigma}%
^{\dagger}f_{i\sigma}\right\rangle $. As they employ the SBMFT the dynamical
fluctuations do not appear explicitly in their Green's Functions (GF), and
they concluded that the multiple peak structure of the specific heat is only
associated with band shape effects. We believe that there is no antagonism
between these two explanations: the dynamical fluctuations modify the bands
(even in the SBMFT through their mean field contributions to the GF), and it
is their spectral densities near the Fermi surface that determine the specific heat.

The dynamical mean field theory\cite{GeorgesKKR} (DMFT), with the help of the
numerical renormalization group (NRG), has been recently employed to study the
PAM\cite{PruschkeBJ} and the Kondo lattice.\cite{CostiM} These works establish
the simultaneous existence of two scales of energy: one high temperature scale
associated with conventional Kondo scattering and a lower temperature scale,
usually called coherence temperature, that is associated with the formation of
quasi-particles with strongly enhanced effective masses.\cite{PruschkeBJ} The
same result has been also obtained by applying the SBMFT in the large-$N$
approach to the Kondo lattice.\cite{BurdinGG} The notation of these
temperature scales is not uniform in the literature, and we shall then use
$T_{0}$ for the lattice coherence temperature,\cite{PruschkeBJ,CostiM} and
introduce $T_{KL}$ for the high temperature scale, leaving $T_{K}$ for the
impurity Kondo temperature.

In the present work we study the thermodynamic properties of the PAM in the
limit of $U\rightarrow\infty$ employing the X-boson method\cite{X-boson2,
FisicaA, X-boson1}in two regimes of the PAM: the heavy fermion Kondo regime
(HF-K) and the local magnetic moment regime (HF-LMM).\cite{Steglich1,
Steglich2} The spurious second order phase transitions of the slave-boson
approach,\cite{Coleman87, X-boson2} at intermediate temperatures and when
$E_{f}<<\mu$, are not present in the X-boson method. This method also includes
explicitly some temporal quantum fluctuations in the GF,\cite{X-boson2} and we
analyze some possible consequences of this difference. We show that the
parameters $T_{0}$ and $T_{KL}$ are determined by the position of the maximum
of $C_{V}$ vs. $T$, and we discuss the physical origin of these maxima. We
show that at low T the entropy per site tends to zero in the HF-K regime,
corresponding to a singlet ground state, while it tends to $k_{B}ln2$ in the
HF-LMM, corresponding to a doublet ground state at each site and opening the
way for the possible formation of an antiferromagnetic order (AF) at low
temperatures ($T<T_{0}$). We show that when the system is in the HF-K regime
and the system goes at low temperatures through the Kondo resonance, there is
a huge increase of the effective mass of the f electrons, associated with the
linear coefficient $\gamma_{o}$ of the specific heat. Our calculations of
$\gamma(T)=C_{v}/T$ qualitatively agree with the experimental results at low
temperatures for some materials classified in the two regimes considered.

\section{ Model and Method}

\label{Sec2}

To study the PAM Hamiltonian in the $U=\infty$ limit, the f-levels with double
occupation are projected out from the space of the local states by employing
the Hubbard $X$ operators,\cite{FFM,Infinite,Ufinito} and we obtain
\begin{align}
H  &  =\sum_{\mathbf{k},\sigma}\varepsilon_{\mathbf{k},\sigma}c_{\mathbf{k}%
,\sigma}^{\dagger}c_{\mathbf{k},\sigma}+\sum_{\sigma,f}\ E_{f,\sigma
}X_{f,\sigma\sigma}\nonumber\\
&  +\sum_{\mathbf{k},\sigma}\left(  V_{f,\mathbf{k},\sigma}X_{f,0\sigma
}^{\dagger}c_{\mathbf{k},\sigma}+V_{f,\mathbf{k},\sigma}^{\ast}c_{\mathbf{k}%
,\sigma}^{\dagger}X_{f,0\sigma}\right)  . \label{Eq.3}%
\end{align}
The first term of the equation represents the Hamiltonian of the conduction
electrons ($c$-electrons), associated with the itinerant electrons ($s$, $p$
and $d$ orbitals). The second term describes the localized $f$ levels and the
last one corresponds to the interaction between the $c$-electrons and the
$f$-electrons via hybridization between the $f$ and $c$ states. This
Hamiltonian can be treated by the X-boson approach\cite{X-boson2, X-boson1,
FisicaA} for the lattice case, and the cumulant Green's function (GF) are then
given by
\begin{equation}
G_{\mathbf{k}{\sigma}}^{ff}(z)=\frac{-D_{\sigma}\left(  z-\varepsilon
_{\mathbf{k}\sigma}\right)  }{\left(  z-\widetilde{E}_{f,\sigma}\right)
\left(  z-\varepsilon_{\mathbf{k}\sigma}\right)  -|V_{\sigma}(\mathbf{k}%
)|^{2}D_{\sigma}}, \label{Eqn.11}%
\end{equation}%
\begin{equation}
G_{\mathbf{k}{\sigma}}^{cc}(z)=\frac{-\left(  z-\widetilde{E}_{f,\sigma
}\right)  }{\left(  z-\widetilde{E}_{f,\sigma}\right)  \left(  z-\varepsilon
_{\mathbf{k}\sigma}\right)  -|V_{\sigma}(\mathbf{k})|^{2}D_{\sigma}},
\label{Eqn.12}%
\end{equation}
and
\begin{equation}
G_{\mathbf{k}{\sigma}}^{fc}(z)=\frac{-\ D_{\sigma}V_{\sigma}(\mathbf{k}%
)}{\left(  z-\widetilde{E}_{f,\sigma}\right)  \left(  z-\varepsilon
_{\mathbf{k}\sigma}\right)  -|V_{\sigma}(\mathbf{k})|^{2}D_{\sigma}},
\label{Eqn.13}%
\end{equation}
where $z=\omega+i\eta$, with $\eta\rightarrow0^{+}$. The correlations appear
in the chain X-boson approach through the quantity $D_{\sigma}=R+n_{f,\sigma}%
$, where $R=\left\langle X_{0,0}\right\rangle $. This simple factor introduces
essential differences with the MFSBT\cite{X-boson2, FisicaA} and these GF can
not be transformed into those of two hybridized bands of uncorrelated
electrons with renormalized parameters, as it is done in the MFSBT. A
renormalized $\widetilde{E_{f}}=E_{f}+\Lambda$ also appears here in the
X-boson method, with $\Lambda$ given by
\begin{align}
\Lambda &  =\frac{1}{N_{s}}\sum_{\mathbf{k},\sigma}\left\vert V_{\sigma
}(\mathbf{k})\right\vert ^{2}\nonumber\\
&  \times\frac{n_{F}\left(  \omega_{\mathbf{k},\sigma}(+)\right)
-n_{F}\left(  \omega_{\mathbf{k},\sigma}(-)\right)  }{\sqrt{\left(
\varepsilon_{\mathbf{k},\sigma}-\widetilde{E}_{f,\sigma}\right)
^{2}+4\left\vert V_{\sigma}(\mathbf{k})\right\vert ^{2}D_{\sigma}}},
\label{Eqn.35a}%
\end{align}
where $n_{F}\left(  x\right)  $ is the Fermi-Dirac distribution and $N_{s}$ is
the number of sites. To simplify the calculations we shall consider a
rectangular conduction band of width $W=2D$, centered at the origin, and a
real hybridization constant $V_{\sigma}(\mathbf{k})=V$. We then obtain
\begin{equation}
\Lambda=\frac{V^{2}}{D}\int_{-D}^{D}d\varepsilon_{\mathbf{k}}\frac
{n_{F}\left(  \omega_{\mathbf{k}}(+)\right)  -n_{F}\left(  \omega_{\mathbf{k}%
}(-)\right)  }{\sqrt{\left(  \varepsilon_{\mathbf{k}}-\widetilde{E}%
_{f}\right)  ^{2}+4V^{2}D_{\sigma}}}, \label{Eqn.36}%
\end{equation}
where the values $\omega_{\mathbf{k},\sigma}(\pm)$ are the poles of the GF,
given by
\begin{align}
&  \omega_{\mathbf{k},\sigma}(\pm)=\frac{1}{2}\left(  \varepsilon
_{\mathbf{k},\sigma}+\widetilde{E}_{f}\right) \nonumber\\
&  \pm\frac{1}{2}\sqrt{\left(  \varepsilon_{\mathbf{k},\sigma}-\widetilde
{E}_{f}\right)  ^{2}+4V^{2}D_{\sigma}}. \label{Eqn.32}%
\end{align}

In the X-boson approach ${D}_{\sigma}=R+n_{f\sigma}$ must be calculated
self-consistently through the minimization of the corresponding thermodynamic
potential with respect to the parameter $R$ . When the total number of
electrons $N_{tot}$, the temperature $T$ and the volume $V_{s}$ are kept
constant one should minimize the Helmholtz free energy $F$, but the same
minimum is obtained by employing the thermodynamic potential $\Omega
=-k_{B}T\ln(\mathcal{Q})$, (where $\mathcal{Q}$ is the Grand Partition
Function) while keeping $T$, $V_{s}$, and the chemical potential $\mu$
constant (this result is easily obtained by employing standard thermodynamic
techniques). In the X-boson method the grand thermodynamic potential
is\cite{X-boson2, FisicaA}
\begin{align}
\Omega &  =\overline{\Omega}_{0}+\frac{-1}{\beta}\sum_{\mathbf{k},\sigma,\pm
}\ln\left[  1+\exp(-\beta\ \omega_{\mathbf{k},\sigma}(\pm))\right] \nonumber\\
&  +N_{s}\ \Lambda(R-1), \label{Eqn.33}%
\end{align}
where
\begin{equation}
\overline{\Omega}_{0}=-\frac{N_{s}}{\beta}\ln\left[  \frac{1+2\exp
(-\beta\widetilde{E}_{f})}{(1+\exp(-\beta\widetilde{E}_{f}))^{2}}\right]  ,
\label{Eqn.34}%
\end{equation}
$\beta=1/k_{B}T$, and $k_{B}$ is the Boltzmann constant. By standard
thermodynamic techniques we find
\begin{equation}
S=-\left(  \frac{\partial F}{\partial T}\right)  _{N_{tot},V_{s}}=-\left(
\frac{\partial\Omega}{\partial T}\right)  _{\mu,V_{s}}, \label{Eqn.40}%
\end{equation}
and
\begin{align}
C_{v}  &  =T\left(  \frac{\partial S}{\partial T}\right)  _{N_{tot},V_{s}%
}=-T\ \left(  \frac{\partial^{2}\Omega}{\ \partial T^{2}}\right)  _{\mu,V_{s}%
}\nonumber\\
&  +T\ \left(  \frac{\partial\mu}{\partial T}\right)  _{N_{tot},V_{s}%
}\ \left(  \frac{\partial N_{tot}}{\partial T}\right)  _{\mu,V_{s}}.
\label{Eqn.41}%
\end{align}
For the rectangular conduction band we find from Eqs. (\ref{Eqn.33}%
,\ref{Eqn.34}) that, in the absence of magnetic field,
\begin{align}
&  -T\ \left(  \frac{\partial^{2}\Omega}{\ \partial T^{2}}\right)  _{\mu
,V_{s}}=-T\ \left(  \frac{\partial^{2}\overline{\Omega}_{0}}{\ \partial T^{2}%
}\right)  _{\mu,V_{s}}\nonumber\\
&  +\frac{k_{B}\hspace{0.1cm}\beta^{2}}{D}\sum_{\ell=\pm}^{2}\int_{-D}%
^{D}dx\hspace{0.1cm}\omega_{\ell}^{2}(x)\hspace{0.1cm}n_{F}\left(
\omega_{\ell}(x)\right)  \left[  1-n_{F}\left(  \omega_{\ell}(x)\right)
\right] \nonumber\\
&  -T\ N_{s}\left(  \frac{\partial^{2}(\Lambda(R-1))}{\partial T^{2}}\right)
_{\mu,V_{s}}, \label{Eqn.43}%
\end{align}
where
\begin{equation}
\omega_\pm(x)=\frac{1}{2}\left(  x+\widetilde{E}_{f}\right)  \pm\frac{1}%
{2}\sqrt{\left(  x-\widetilde{E}_{f}\right)  ^{2}+4\left\vert V\right\vert
^{2}D_{\sigma}} \label{Eqn.44}%
\end{equation}
and
\begin{align}
&  -T\ \left(  \frac{\partial^{2}\overline{\Omega}_{0}}{\ \partial T^{2}%
}\right)  _{\mu,V_{s}}=-2N_{s}\ k_{B}\hspace{0.1cm}\beta^{2}\widetilde{E}%
_{f}^{2}\hspace{0.1cm}\exp(\beta\widetilde{E}_{f})\nonumber\\
&  \qquad\times\frac{\lbrack3+2\exp(\beta\widetilde{E}_{f})]}{\left[
\exp(\beta\widetilde{E}_{f})+2\right]  ^{2}\left[  \exp(\beta\widetilde{E}%
_{f})+1\right]  ^{2}}. \label{Eqn.45}%
\end{align}

We have calculated $T\ \left(  \partial\mu/\partial T\right)  _{N_{tot},V_{s}%
}\ \left(  \partial N_{tot}/\partial T\right)  _{\mu,V_{s}}$ numerically.

\section{Results and Discussion}

\label{Sec3}

A schematic classification for typical Ce-based compounds was given earlier by
Varma\cite{Varma85} and reintroduced by Steglich et. al.\cite{Steglich1,
Steglich2} in terms of the dimensionless coupling constant for the exchange
between the local $f$ spin and the conduction-electron spins, $g=N_{F}\ |J|$,
where $N_{F}$ is the conduction-band density of states at the Fermi energy and
$J$ is connected to the parameters of the PAM via a Schrieffer-Wolff
transformation,\cite{SchriefferWolff} $J=|2V^{2}/\left(  {\ E}_{f}-\mu\right)
|$ (for $U\rightarrow\infty$). The behavior of these compounds can then be
qualitatively organized through the parameter%

\begin{equation}
g=2V^{2}\rho_{c}\left(  \mu\right)  /|E_{f}-\mu|, \label{g}%
\end{equation}

\noindent where $\rho_{c}\left(  \mu\right)  =1/2D$ is independent of $\mu$ in
our case, $E_{f}$ is the bare localized $f$ level and we employ the chemical
potential $\mu$ obtained in the self-consistent X-boson calculation for the
total number of particles per site $n_{tot}=2(n_{f,\sigma}+n_{c,\sigma})$ at
the very low temperature $T=0.0001D$.

When $g>1$, the compound under consideration is in the intermediate valence
(IV) region, while for $g<1$ it is in the heavy fermion Kondo regime (HF-K).
There exists a critical value $g_{c}$ at which the Kondo and the RKKY
interactions have the same strength, and non Fermi-liquid (NFL) effects have
been postulated for systems with $g=g_{c}$. For $g_{c}<g<1$ the magnetic local
moments are quenched at very low temperatures and the system presents a Fermi
liquid behavior, while for $g<g_{c}$ we have the local magnetic moment regime
(HF-LMM). We point out that the parameter $g$ classifies the regimes of the
PAM only in a qualitative way. Finally, as in its present form the X-boson
approach includes hybridization effects only to second order in $V$ and the
self-energy does not depend on the wave vector,\cite{X-boson2} the RKKY
effects are not present and we cannot discuss non Fermi-liquid behavior within
the present approximation, nor find the value of $g_{c}$.

\subsection{Results for the HF- regime}

\begin{figure}[th]
\includegraphics[clip,width=0.40\textwidth,
height=0.35\textheight,angle=-90.]{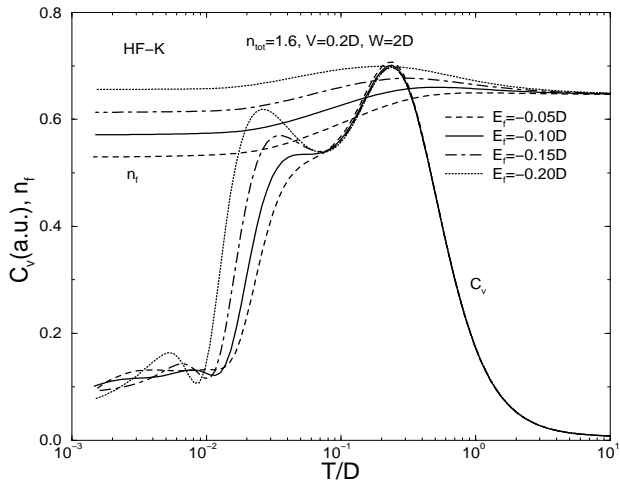}\caption{Specific
heat $C_{V}$ and occupation of the localized $f$ levels $n_{f}$
vs. temperature $T$, for the HF-K regime. Notice that the
intensity of the two maxima of $C_{V}$ at
lower $T$ first decreases and then disappears when the number $n_{c}%
=1.6-n_{f}$ of conduction electrons increases, while their positions $T_{0}$
and $T_{KL}$ shift to the right.}%
\label{fig1_CvL}%
\end{figure}

\begin{figure}[th]
\includegraphics[clip,width=0.40\textwidth,
height=0.35\textheight,angle=-90.]{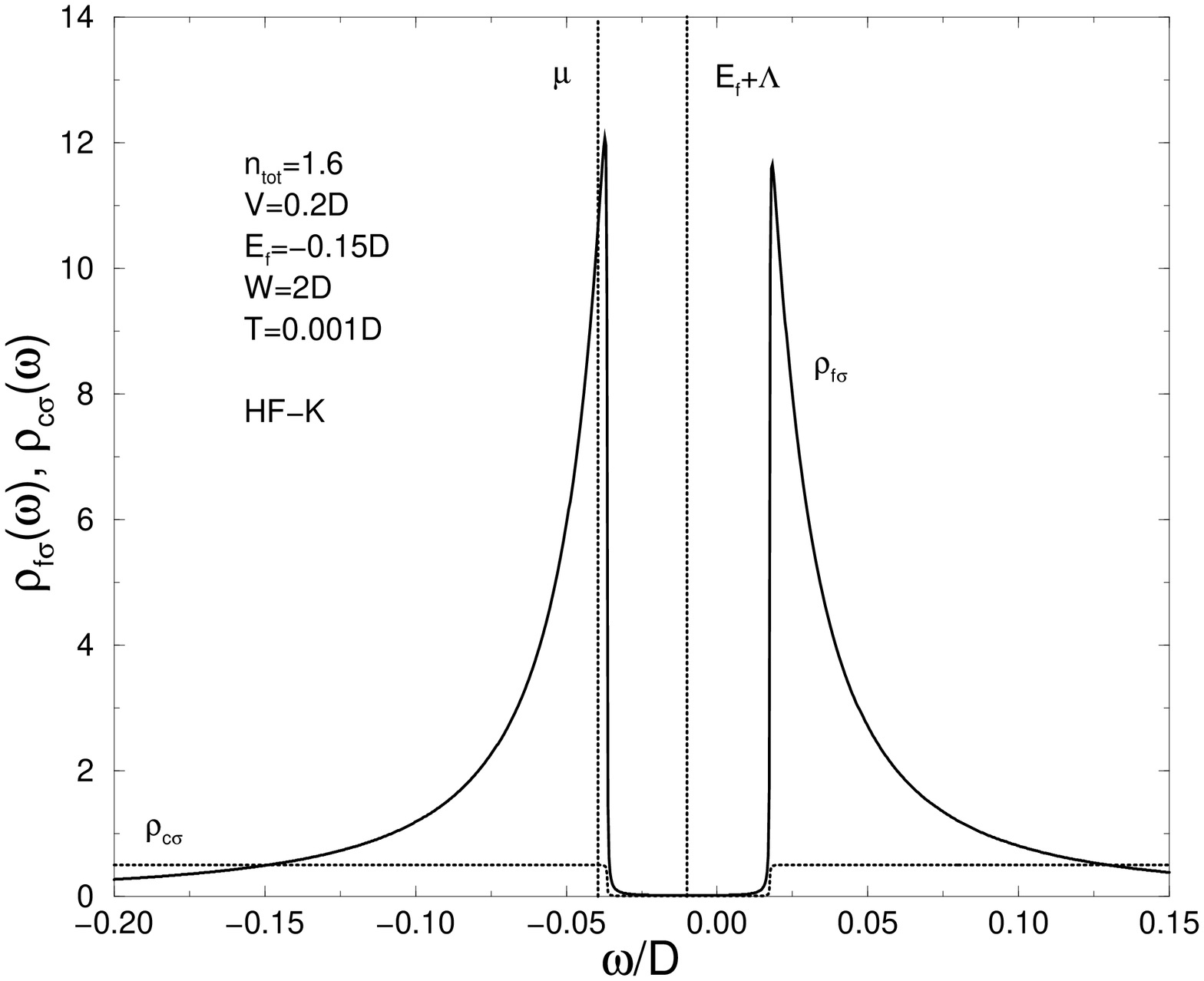 }\caption{The
$f$ and $c$ electronic density of states $\rho_{f,\sigma}(\omega)$
and $\rho_{c,\sigma }(\omega)$ vs. $\omega$ at low temperatures,
for the HF-K regime. The position of the renormalized energy level
$\tilde{E}=E_{f}+\Lambda$ and of the chemical potential $\mu$ are
shown. Similar results were obtained for other set of
parameters in the same regime.}%
\label{fig2_CvL}%
\end{figure}

In Fig. \ref{fig1_CvL} we show the specific heat $C_{v}$ and the occupation
$n_{f}$ of the localized level as a function of the temperature $T$ for
several sets of parameters that describe a HF-Kondo situation (HF-K):
hybridization $V=0.2D$, total number of particles per lattice site
$n_{tot}=1.6$ and bandwidth $W=2D$; the bare localized energy levels $E_{f}$
employed and the corresponding Steglich's parameters\cite{Steglich1,
Steglich2,Note1} $g$ (for $T=0.0001D$) are $E_{f}=-0.05D,$ $g=0.426$;
$E_{f}=-0.10D,$ $g=0.392$; $E_{f}=-0.15D,$ $g=0.365$ and $E_{f}=-0.20D,$
$g=0.342$.
In Fig. \ref{fig2_CvL} we plot both the $f$ and $c$ spectral densities for
$E_{f}=-0.15D$, showing the high density of states $f$ at the chemical
potential for that set of parameters.

\begin{figure}[th]
\includegraphics[clip,width=0.40\textwidth,
height=0.35\textheight,angle=-90.]{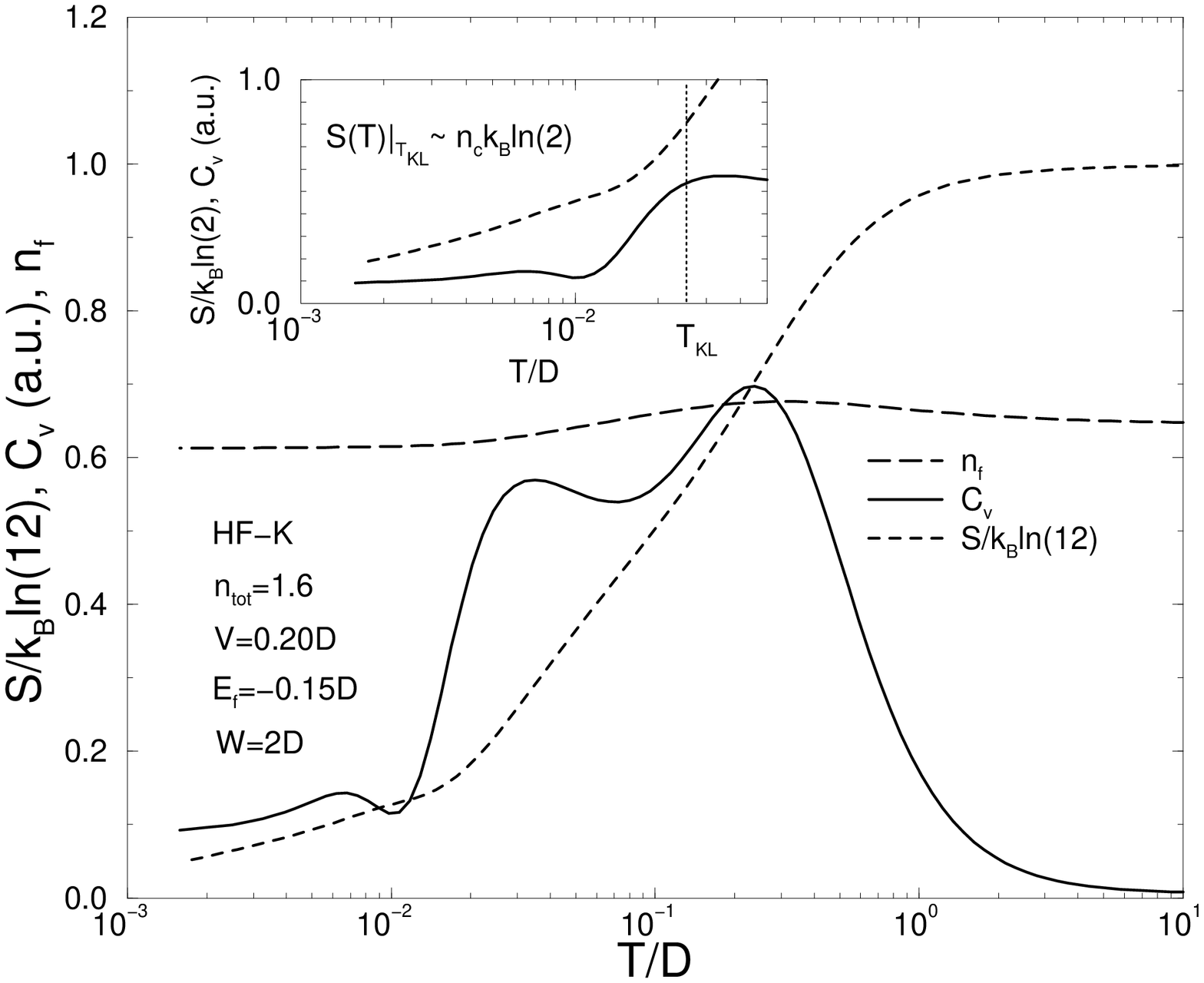 }\caption{Entropy
$S$, Specific heat $C_{v}$ and total occupation of the localized
levels $n_{f}$ vs. $T$, for the same set of parameters of Fig.
\ref{fig2_CvL}. The behavior of the curves
is similar for other parameters corresponding to the same HF-K regime.}%
\label{fig3_CvL}%
\end{figure}

In Fig. \ref{fig3_CvL} we plot the specific heat, the entropy and the total
occupation of the $f$ level as a function of $T$ for the same set of
parameters of Fig. \ref{fig2_CvL}. We can see that the entropy $S$ per site
tends to zero at low temperatures, and we conclude that the system goes to the
Kondo singlet ground-state, while the huge value of density of states $f$ at
the chemical potential $\mu$ (see figure \ref{fig2_CvL}) signals the
simultaneous appearance of the Kondo resonance. The inset shows, in agreement
with reference \onlinecite{BurdinGG}, that an entropy approximately equal to
$n_{c}\ k_{B}\ ln(2)$ is lost below $T_{KL}$.

\begin{figure}[th]
\includegraphics[clip,width=0.40\textwidth,
height=0.35\textheight,angle=-90.]{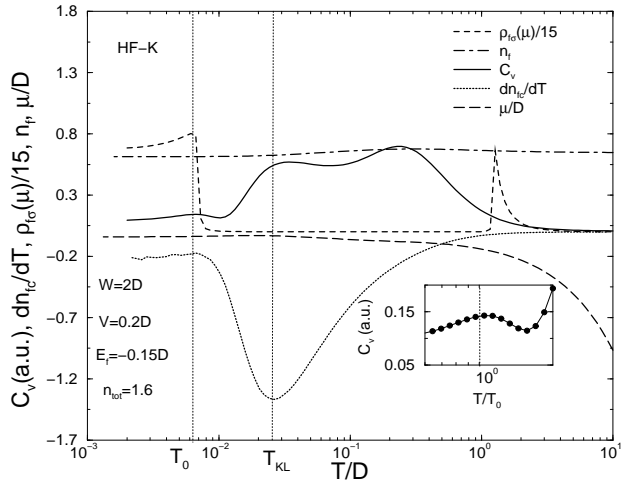}\caption{Specific
heat $C_{v}$, chemical potential $\mu$, density of states
$\rho_{f,\sigma}(\mu)$ of the $f$-electrons at $\mu$, $\left(
dn_{fc,\sigma}/dT\right)  $, and $n_{f}$ vs. $T$, for the same set
of parameters given in Fig. \ref{fig1_CvL} but with
$E_{f}=-0.15D$. We indicate the temperatures $T_{0}$
\cite{Portuges} and $T_{KL}$.\cite{Lacroix} The inset shows the
little peak in $C_{v}$ at $T=T_{0}$. The behavior for other sets
of parameters corresponding to the same
regime is similar.}%
\label{fig4_CvL}%
\end{figure}

\begin{figure}[th]
\includegraphics[clip,width=0.40\textwidth,
height=0.35\textheight,angle=-90.]{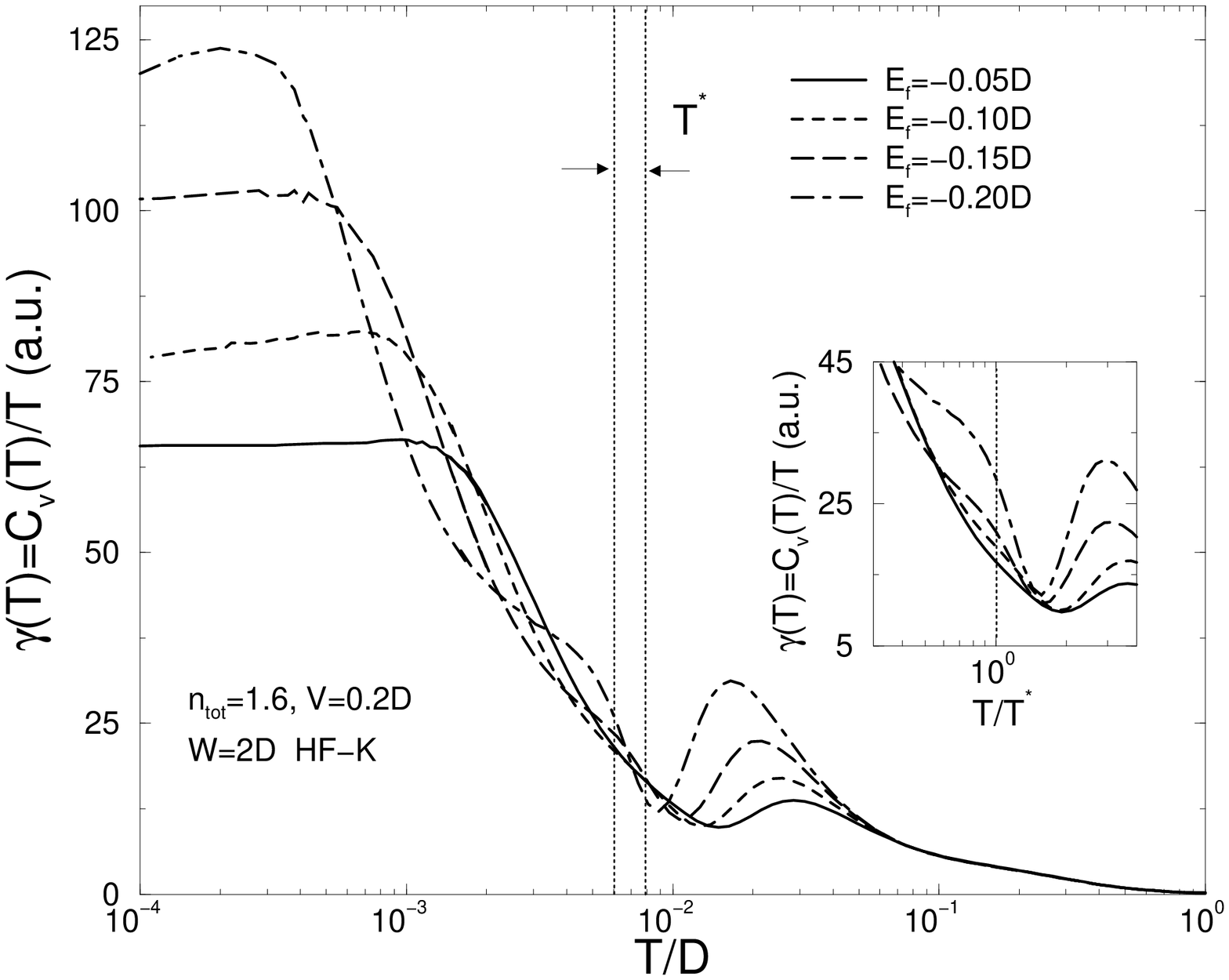
}\caption{Sommerfeld's coefficient $\gamma(T)=C_{v}(T)/T$ vs. $T$
at very low temperatures for the HF-K regime. We have
$\gamma(T)|_{T=0.0}=\gamma _{o}\propto m^{\ast}$, where $m^{\ast}$
is the effective mass of the quasi-particles. The two vertical
lines delimit the interval of values of $T_{0}$ corresponding to
the four different curves. In the inset one sees an anomaly in
$\gamma(T)$ at $T=T_{0}$ and verifies that $T^{\gamma}<<T_{0}$ for
all the values of $E_{f}$.}%
\label{fig5_CvL}%
\end{figure}

\begin{figure}[th]
\includegraphics[clip,width=0.40\textwidth,
height=0.35\textheight,angle=-90.]{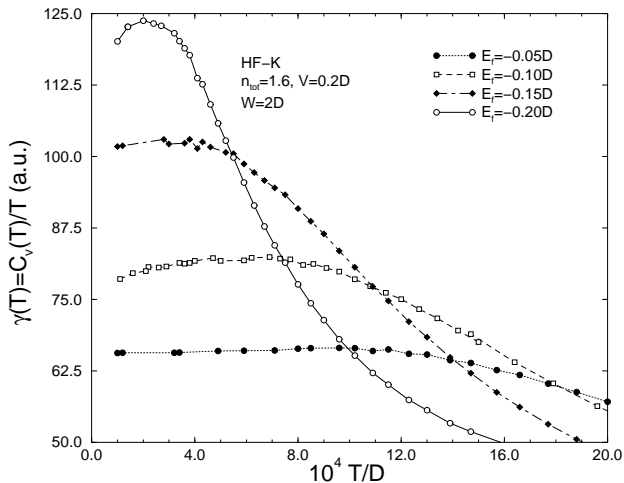}\caption{Sommerfeld's
coefficient $\gamma(T)=C_{v}(T)/T$ vs. $T$ for the HF-K regime, at
very low
temperatures ($T\sim T^{\gamma}$).}%
\label{fig6_CvL}%
\end{figure}

At high temperatures the system behaves like a collection of independent $f$
and $c$ electrons, because both $V/k_{B}T$ and $D/k_{B}T$ tend to zero like
$1/T$. For a system with a chemical potential that remains finite when
$T\rightarrow\infty$ the entropy per site would tend to $k_{B}\ ln(12)$,
corresponding to the dimensionality of the purely local Hilbert space, but
this result is not true any more when we keep fixed the total number of
electrons per site $n_{tot}$, as discussed in more detail in Appendix
\ref{ApA}. The value of the entropy when $n_{tot}=1.6$ (the value employed in
figures \ref{fig1_CvL}-\ref{fig6_CvL}) is equal to $0.998765\ k_{B}\ ln12$ and
is indistinguishable from $k_{B}\ ln(12)$ in the figures.

In the Fig. \ref{fig4_CvL} we plot the specific heat $C_{v}$, the chemical
potential $\mu$,\ the temperature derivative of the parameter $\left\langle
c_{i\sigma}^{\dagger}f_{i\sigma}\right\rangle \equiv n_{fc,\sigma}$, the
occupation number $n_{f}$ of the $f$ levels, and the value at $\mu$ of the $f$
density of states $\rho_{f,\sigma}(\mu)$, as a function of $T$ for a system in
the HF-K regime. The average $\left\langle c_{i\sigma}^{\dagger}f_{i\sigma
}\right\rangle \equiv n_{fc,\sigma}$ is associated with the GF $G_{\mathbf{k}%
{\sigma}}^{fc}(z_{n})$, and it is a measure of the mixing of the $c$ and $f$
states in the system. We associate the temperatures $T_{0}$ and $T_{KL}$
respectively to the maximum and the minimum of the $\left(  dn_{fc,\sigma
}/dT\right)  $ \emph{(they are named }$T^{\ast}$\emph{ in Peres
et.al.\cite{Portuges} and }$T_{K}$\emph{ in Bernhard et.al.\cite{Lacroix}
respectively)}. They are both shown in the figure, where it is clear that
$T_{0}$ coincides with the first peak of $C_{V}$ and $T_{KL}$ with the second
one. Only one extreme was observed in previous works.\cite{Lacroix, Portuges}

The chemical potential $\mu$ changes with temperature because we keep the
total number of particles constant, and in our calculations it shifts into the
hybridization gap at intermediate temperatures. The density of states at the
chemical potential $\rho_{f,\sigma}(\mu)$ has a very large value at low $T$,
but at $T\sim T_{0}$ this quantity shows a drastic change with increasing $T$,
taking a value close to zero when $\mu$ enters the gap, as shown in Fig.
\ref{fig4_CvL}. The moderate increase in the average electronic energy that
occurs in this process is the origin of the first peak of $C_{v}$.

It seems reasonable to use $T_{0}$\ as a measure of the Kondo lattice
temperature, because crossing $T_{0}$ (with decreasing $T$) we find that
$\rho_{f,\sigma}(\mu)$ changes from very small to very large values, and at
the same time the entropy takes small values, tending to zero when
$T\rightarrow0$ (cf. Fig. \ref{fig3_CvL}). These two properties characterize
respectively the appearance of the Kondo peak and the formation of a singlet
by the quenching of the magnetic moment of the $f$ electrons by the cloud of
$c$ electrons, and our choice of $T_{0}$ as the crossover parameter seems reasonable.

The temperatures $T_{0}$ and $T_{KL}$ coincide in our treatment with the first
two peaks of $C_{V}$, and correspond to the two energy scales that appear in
the DMFT treatment of both the PAM\cite{PruschkeBJ} and of the Kondo
lattice.\cite{CostiM} The lattice coherence scale is associated in the first
reference\cite{PruschkeBJ} to the inverse effective mass at the Fermi surface,
which in our treatment is proportional to the\ corresponding density of
states: as shown in Fig. \ref{fig4_CvL} this quantity develops a very large
increase when $T$ decreases below $T_{0}$. This same scale is also associated
in the second reference\cite{CostiM} to the onset of Fermi-liquid coherence,
and can be determined from various dynamical quantities including the
$f$-electron Kondo resonance.\cite{Costi} It should be remarked that the
drastic change that $\rho_{f\sigma}(\mu)$ suffers when $T$ crosses $T_{0}$
(cf. Fig. \ref{fig4_CvL}), is consistent with the result reported in Section
3.5 of reference,\cite{CostiM} that the position of the peak of the resonance,
\textquotedblleft relative to the chemical potential shifts to below the Fermi
level on a scale $\lesssim T_{0}$ with increasing
temperature.\textquotedblright

The second peak at $T=T_{KL}$ corresponds to a second and larger maximum of
$\left\vert d\left\langle c_{i\sigma}^{\dagger}f_{i\sigma}\right\rangle
/dT\right\vert $, corresponding to the maximum rate of change with $T$ of the
mixing between $f$ and $c$ electrons. Even if the GF do not have explicit
dynamic fluctuations, the changes in $\left\vert \left\langle c_{i\sigma
}^{\dagger}f_{i\sigma}\right\rangle \right\vert $ measure a mean-field average
of the charge fluctuations, and the peak is due to the energy necessary to
change the average occupation of the different types of electrons. The
crossover parameter $T_{KL}$ has to be used when $\left\vert d\left\langle
c_{i\sigma}^{\dagger}f_{i\sigma}\right\rangle /dT\right\vert $ shows a single
extreme, as it happens in the half filled symmetric PAM,\cite{Lacroix} because
$\mu$ is then fixed in the middle of the gap and the changes in population
that originated $T_{0}$ do not occur.

The temperature $T_{KL}$ corresponds to the second energy scale in
both DMFT works,\cite{PruschkeBJ,CostiM} and its corresponding
energy is related to the transitions between the states of the
pseudo gap that usually appears in that treatment. Our method only
gives a real self energy, and we always obtain a real gap rather
than a pseudo gap, but the meaning of $T_{KL}$ remains the same.
The energy associated to $T_{KL}$ has essentially the same origin
as the one corresponding to the Kondo impurity $T_{K}$, describing
the conventional incoherent Kondo scattering at higher
temperatures.\cite{PruschkeBJ} In agreement with the results of
several authors,\cite{PruschkeBJ,CostiM,BurdinGG}  $T_{KL}$
decreases when the number $n_{c}$ of conduction electrons
decreases.

Two scales are also found in the large-$N$ SBMFT of the Kondo
lattice,\cite{BurdinGG} characterized by the two maxima of $C_{V}$ vs. $T$
(though the second maximum is replaced in their Fig. 3b by a discontinuity at
the phase transition that appears in the SBMFT at $T_{KL}$). Although both the
model and the treatment are different, the $C_{V}$ maxima at $T_{0}$ in our
Fig. \ref{fig1_CvL} have the same trend they show in reference
\onlinecite{BurdinGG}, namely that they decrease and finally disappear when
the number $n_{c}$ of conduction electrons increases.

Finally the third peak of $C_{V}$ is located in a region corresponding to
large $T$, and it measures the energy necessary to take the electrons from the
low $T$ distribution into the high temperature limit distribution, that
depends on $n_{tot}$ (it would be the uniform distribution for $n_{tot}=5/3$,
cf. Appendix \ref{ApA}). As could be expected, the position of the high
temperature peak is independent of the value of $E_{f}$, as it is shown in
figure \ref{fig1_CvL}.

In Fig. \ref{fig5_CvL} we plot the coefficient $\gamma(T)=C_{V}(T)/T$ vs. $T$
for several values of $E_{f}$ corresponding to the HF-K regime. At low $T$ the
coefficient $\gamma(T)$ is proportional to the density of states at the
chemical potential $\mu$, and in simple cases this is proportional to the
effective mass $m^{\ast}$ of the associated quasi-particles. We see in that
figure that the value of $\gamma(T)|_{T=0.0}=\gamma_{o}$ at very low
temperatures \emph{has a large increase} when $E_{f}$ changes from
$E_{f}=-0.05D$ to $E_{f}=-0.20D$, while the Steglich's parameter respectively
decreases from $g=0.426$ to $g=0.342$. This shows that the quasi-particles
become heavier as the system evolves in the direction of the Kondo regime. The
maximum of $\gamma(T)$ is reached at a value $T^{\gamma}$ that satisfies
$T^{\gamma}<<{T_{0}}$
. The inset of Fig. \ref{fig5_CvL} shows $\gamma(T)$ vs. $T/T_{0}$ for
$T/T_{0}\sim1$, and indicates an evident anomaly in $\gamma(T)$ at $T={T}_{0}$
for all the values of $E_{f}$ in the HF-K parameter region. The general form
of our results agrees with the experimental measurements for HF
systems,\cite{Steglich1, Amato87b, Steglich85, Benoit, Bredl, Rebelsky,Andres}
as well as with those of a recent theoretical work.\cite{Chinos}

In Fig. \ref{fig6_CvL} we present the same results of Fig. \ref{fig5_CvL} for
very low temperatures showing the maximum of $\gamma(T)$ in detail. The
results are qualitatively similar to the experimental measurements for
different materials,\cite{Steglich1} like $CeAl_{3}$\cite{Steglich1, Benoit,
Bredl, Steglich85, Andres} and $CePtSi.$\cite{Steglich1, Rebelsky} It is
important to emphasize that the best qualitative agreement with the
experimental results is given by the set of parameters that corresponds to the
highest values of $\rho_{f,\sigma}(\mu)$ at very low temperatures, pointing
out the strong HF character of these compounds.

\subsection{Results for the HF-LMM regime}

\begin{figure}[th]
\includegraphics[clip,width=0.40\textwidth,
height=0.35\textheight,angle=-90.]{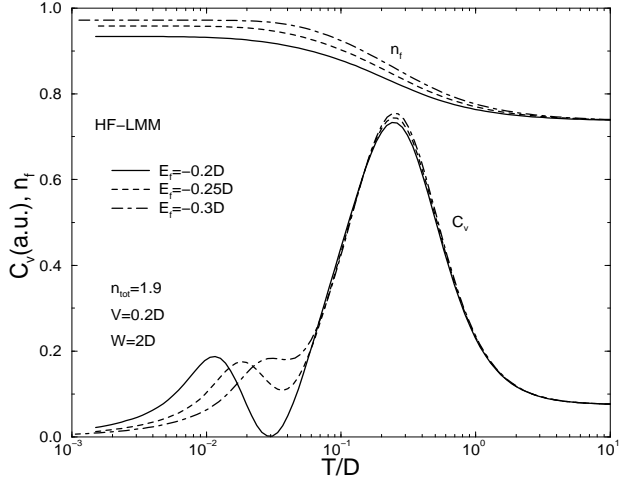}\caption{Specific
heat $C_{v}$ and occupation on the localized $f$ levels $n_{f}$
vs. temperature $T$, for
the HF-LMM regime.}%
\label{fig7_CvL}%
\end{figure}

Now we consider several sets of parameters associated with the heavy Fermion
local magnetic regime (HF-LMM), defined as before through the Steglich's
criterion $g<g_{c}<1$,\cite{Steglich1, Steglich2} but characterized in our
treatment by its properties.\cite{Note1} Our calculations in this regime show
that the density of $f$ states at the chemical potential $\rho_{f,\sigma}%
(\mu)$ for $T\rightarrow0$ has low values, that the specific heat has a two
peak structure, and that the occupation number $n_{f}$ of localized states is
close to $1$. In Fig. \ref{fig7_CvL} we plot the same properties given in Fig.
\ref{fig1_CvL} for the following set of parameters, that correspond to the
HF-LMM regime: hybridization $V=0.2D$, total number of particles per site
$n_{tot}=1.9$, bandwidth $W=2D$, and localized energy level given by
$E_{f}=-0.20D$, $E_{f}=-0.25D$, and $E_{f}=-0.30D$, corresponding to
Steglich's parameters $g=0.216$,$\ g=0.191$, and $g=0.164$ respectively.

\begin{figure}[th]
\includegraphics[clip,width=0.40\textwidth,
height=0.35\textheight,angle=-90.]{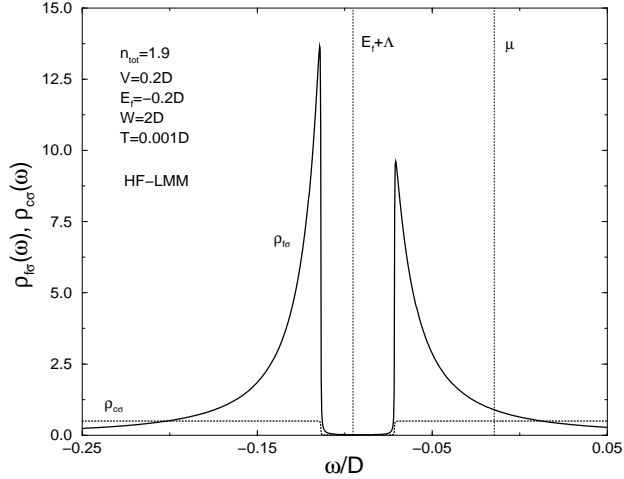}\caption{Density of states
$\rho_{f,\sigma}(\omega)$ and $\rho_{c,\sigma}(\omega)$ of $f$ and $c$
electrons as a function of $\omega$ at low temperature, for parameters
corresponding to the HF-LMM regime. The position of the renormalized energy
level $\tilde{E}=E_{f}+\Lambda$ and of the chemical potential $\mu$ are shown.
Other sets of parameters in the same regime show a similar behavior.}%
\label{fig8_CvL}%
\end{figure}

In Fig. \ref{fig8_CvL} we plot both the $f$ and $c$ spectral densities for
$E_{f}=-0.20D$ and $T=0.001D$, showing the rather small value of the spectral
density $\rho_{f,\sigma}(\mu)$ at the chemical potential $\mu$. Notice that
in this case (HF-LMM regime) the renormalized energy $\widetilde{E}_{f}%
=E_{f}+\Lambda$ is well below the chemical potential $\mu$.

\begin{figure}[th]
\includegraphics[clip,width=0.40\textwidth,
height=0.35\textheight,angle=-90.]{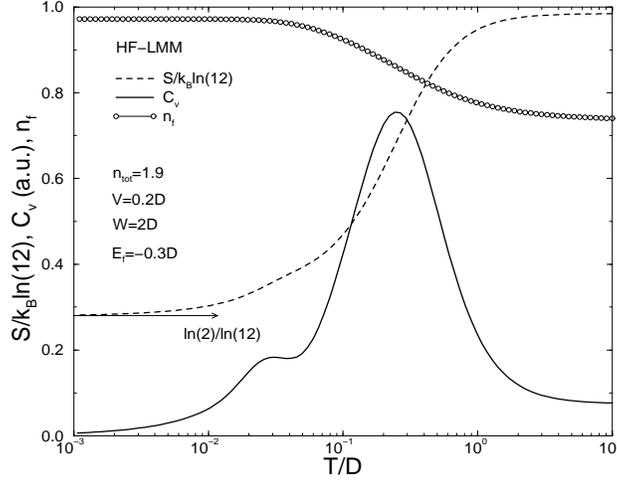}\caption{Entropy
$S$, specific heat $C_{v}$ and charge at the localized levels
$n_{f}$ vs. $T$ for a set of parameters corresponding to the
HF-LMM regime. Other sets of parameters in the
same regime show a similar behavior.}%
\label{fig9_CvL}%
\end{figure}

The Fig. \ref{fig9_CvL} is similar to Fig. \ref{fig3_CvL}, but for the system
in the HF-LMM regime. At very low temperatures the entropy $S$ per site
\ tends to $k_{B}ln(2)$, pointing to the transformation of the singlet of the
HF-K regime into a ground state consisting of a doublet at each site, that
could be attributed to a spin $1/2$ at each site. Like in the HF-K regime, the
entropy $S$ goes to a high temperature limit that depends only on $n_{tot}$,
as discussed in Appendix \ref{ApA}.

In Fig. \ref{fig10_CvL} we present the same results of Fig. \ref{fig4_CvL} for
the HF-LMM regime. In this regime the $\rho_{f,\sigma}(\mu)$ goes to zero only
in the limit of very high temperatures, because $\mu$ is always in the tail of
the density of states (see figure \ref{fig8_CvL}) and does not cross through
the gap, as it does in Fig. \ref{fig4_CvL}. As a consequence, the peak
corresponding to the one at the lowest $T$ in figures \ref{fig1_CvL},
\ref{fig3_CvL} and \ref{fig4_CvL} is absent in this regime, and the first
specific heat peak that appears is physically equivalent to the one at
$T_{KL}$ in the HF-K regime, i.e.: it corresponds to a redistribution of
states into higher energies. This $C_{V}$ peak is less pronounced than in the
HF-K region, because many of \ the low energy states are already occupied at
small $T$ because of the larger value of $N_{t}$. The second peak of $C_{V}$
is now associated with the high temperature limit, and the same arguments
employed in the HF-K region apply here; \ in particular its position is
independent of the value of $E_{f}$ (cf. Fig. \ref{fig7_CvL}).

\begin{figure}[th]
\includegraphics[clip,width=0.40\textwidth,
height=0.35\textheight,angle=-90.]{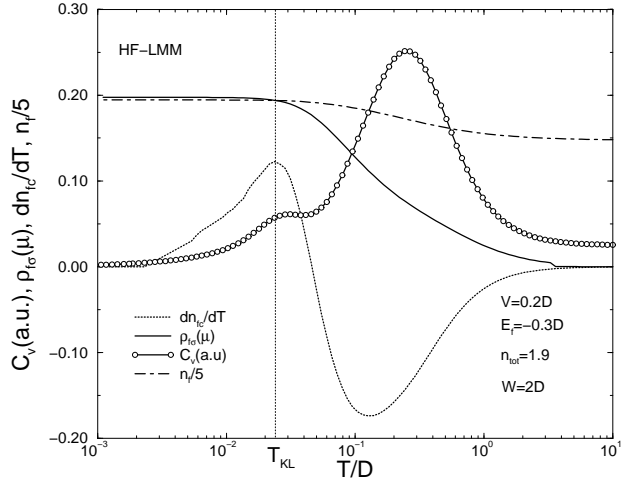}\caption{Specific heat
$C_{V}$, density of local states $\rho_{f,\sigma}(\mu)$ on the
chemical potential $\mu $, $dn_{fc}/dT$ and $n_{f}$ vs. $T$, for a
set of parameters corresponding to the HF-LMM regime. Only the
maximum of $dn_{fc}/dT$ coincides with a maximum of $C_{V}$ at
$T_{KL}$. Other sets of parameters in the same regime show a
similar behavior.}%
\label{fig10_CvL}%
\end{figure}

\begin{figure}[th]
\includegraphics[clip,width=0.40\textwidth,
height=0.35\textheight,angle=-90.]{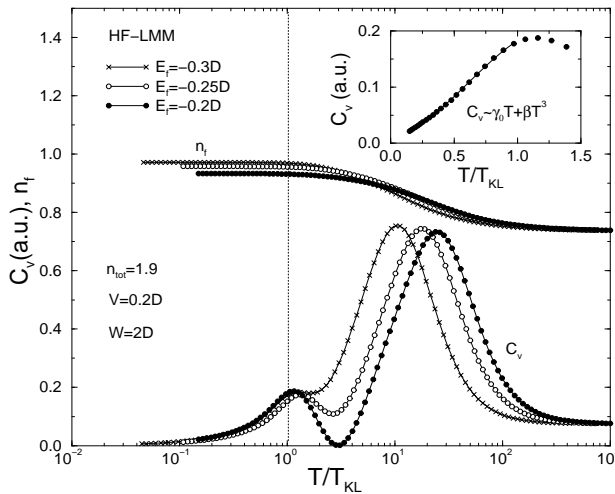}\caption{Specific
heat and occupation numbers of $f$ electrons vs. $T/T_{KL}$ in the
HF-LMM regime. The inset presents the $T/T_{KL}<1$ region, showing
a cubic contribution $C_{v}\propto\gamma_{o}\left(
T/T_{KL}\right)  +\beta\left(  T/T_{KL}\right)
^{3}$.}%
\label{fig11_CvL}%
\end{figure}

In Fig. \ref{fig11_CvL} we present the same results of Fig. \ref{fig7_CvL}
(HF-LMM regime) but with the temperature scaled by $T_{KL}$. All the first
peaks collapse at $T=T_{KL}$ by construction, and the inset shows a universal
cubic dependence of $C_{v}$ vs. $T$ at low temperatures. The same dependence
was obtained by Tesanovic and Valls\cite{TesanovicV} by starting from the
slave boson mean field treatment and considering the interaction between heavy
fermions, mediated by fluctuations of the slave bosons. As discussed after
equation (A11) in reference \onlinecite{X-boson2}, the X-boson method retains
some time dynamics that is absent in the slave boson method, and this might
lead to the same effect obtained by Tesanovic and Valls.

Kim and Stewart\cite{KimS} conclude from their experimental results that, in
some heavy Fermion compounds like \textrm{CeCu}$_{6}$, a significant
contribution to the $C_{V}$ comes from magnetic correlations. We are not
explicitly including magnetic interactions into the model Hamiltonian, and as
discussed before, our treatment neither includes the RKKY interaction.
Therefore, we can not attribute this $T^{3}$ dependence of $C_{V}$ to
intersite magnetic interactions, but it would correspond to Sommerfeld's
second tem in the $C_{V}$ temperature expansion.

\begin{figure}[th]
\includegraphics[clip,width=0.40\textwidth,
height=0.35\textheight,angle=-90.]{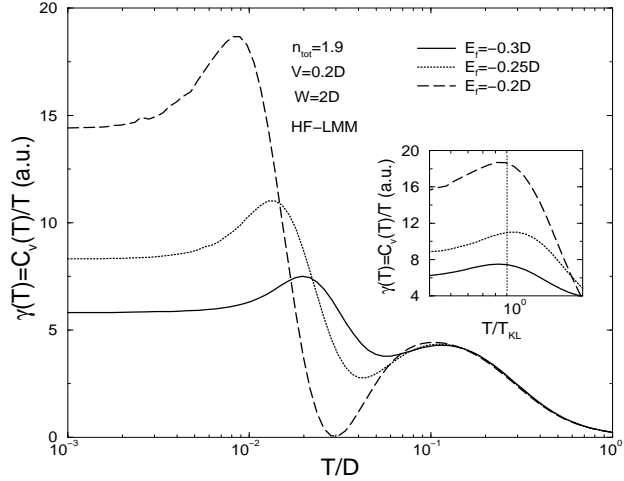}\caption{Sommerfeld's
coefficient $\gamma(T)=C_{v}(T)/T$ vs. $T$ in the HF-LMM regime at very low
temperatures. The coefficient $\gamma(T)|_{T=0.0}=\gamma_{o}\propto m^{\ast}$,
where $m^{\ast}$ is the reduced mass of the quasi-particles.}%
\label{fig13_CvL}%
\end{figure}

\begin{figure}[th]
\includegraphics[clip,width=0.40\textwidth,
height=0.35\textheight,angle=-90.]{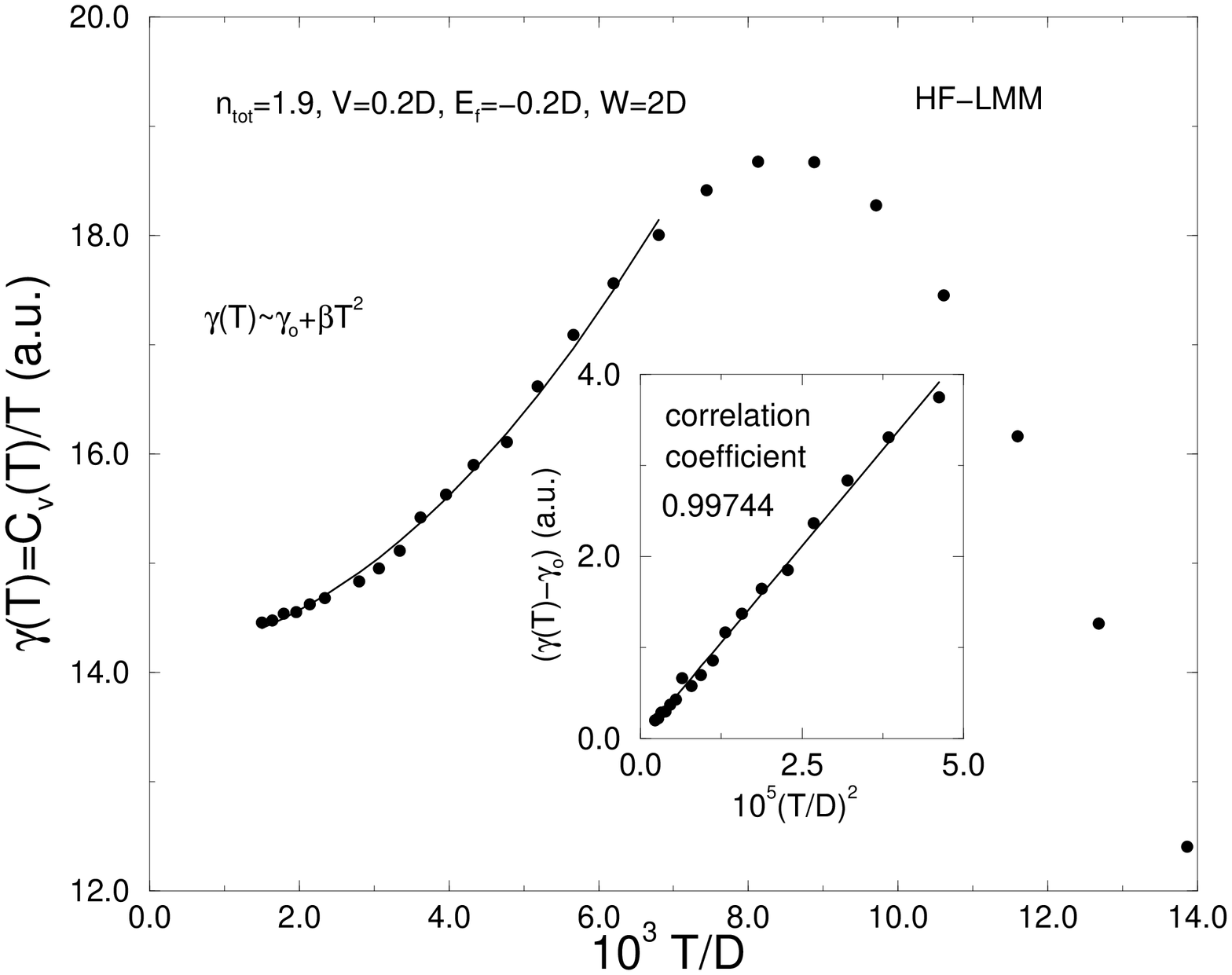
}\caption{Sommerfeld's coefficient $\gamma(T)=C_{v}(T)/T$ vs. $T$
in the HF-LMM regime at very low temperatures. The inset shows the
linear behavior of the $\left(  \gamma(T)-\gamma _{o}\right)  $
vs. $T^{2}$ associated with Sommerfeld's second term in the $T$
expansion of $C_{V}$.}%
\label{fig14_CvL}%
\end{figure}

In Fig. \ref{fig13_CvL} we show the coefficient $\gamma(T)=C_{v}/T$ \ vs. $T$
for the HF-LMM regime, and in the inset we show $\gamma(T)=C_{v}(T)/T$ vs
$T/T_{KL}$. The reduction of the effective masses $m^{\ast}\propto\gamma_{o}$
corresponds to the crossover from the HF-LMM regime to Steglich's stable
moment regime (SMR ) corresponding\cite{Steglich1, Steglich2} to $g<<1$. The
reduction of the effective mass is again associated to the decrease of
$\rho_{f,\sigma}(\mu)$ at very low temperatures ($T<T_{KL}$).

In Fig. \ref{fig14_CvL} we present in a linear scale the same results of Fig.
\ref{fig13_CvL} at $T\sim T_{KL}$, but only for $E_{f}=-0.20D$. The inset
shows the linear behavior of $\left(  \gamma(T)-\gamma_{o}\right)  $ vs.
$T^{2}$, corresponding to the cubic dependence of $C_{V}$ discussed above.

\subsubsection{Discussion}

It is clear that our method does not give the Kondo peak nor the quenching of
the magnetic moment in the HF-LMM region. This is the region where the RKKY
interactions override the Kondo effect, and one is not quite clear of what
should be the exact paramagnetic solution in that region. The Kondo resonance
is observed in some theoretical calculations of the paramagnetic region with a
moderately small value of Steglich's parameter $g$\cite{Steglich1,Steglich2},
like in Fig. 3f of reference \onlinecite{Chinos}, but we are not certain that
these calculations correspond to the HF-LMM region. We must also consider that
we have not included explicit magnetic interactions into the Hamiltonian, and
also that there is no RKKY interaction because only second order cumulants are
employed in the basic CHA expansion, as shown by the corresponding diagrams
(see e.g. Fig. 1 or reference \onlinecite{X-boson2}).

The absence of fourth and higher order cumulants in the CHA leads to a
disturbing question: how do we obtain any Kondo resonance at all in the HF-K
region, if apparently there are no spin flip processes in our calculation? To
answer this question we notice that although the X-boson method starts from an
expansion without spin flips, it is followed by a minimization of the free
energy with respect to $R=\left\langle X_{00}\right\rangle $, with the
completeness as a subsidiary condition (cf. Eq. (4) of reference
\onlinecite{X-boson2}). In this procedure the original solution is modified so
that it gives the best possible physical solution (as regards to the free
energy $F$) compatible with the general form of the CHA. The shifting of the
localized energy $E_{f}$ to the renormalized value $\widetilde{E_{f}}%
=E_{f}+\Lambda$ optimizes the free energy $F$ giving at the same time a peak
near the chemical potential $\mu$, that takes the place of the Kondo resonance
when parameters in the HF-K region are used. This result is not achieved in
the HF-LMM region because the required shift $E_{f}$ would not minimize $F$,
and one is then left with an unquenched doublet without the corresponding
spectral density increase around $\mu$. A complementary explanation is that
the constrained minimization would correspond to adding to the CHA expansion
an extra number of undefined higher order diagrams, that would result in the
formation of the Kondo resonance and of the quenching of the magnetic moment
by the conduction electrons in the HF-K region.

We should point out that the same type of arguments presented here would also
apply to the SBMFT, because after making the mean field approximation of the
slave boson operator, one is left with two hybridized bands without any
explicit correlation, and it is again the free energy restrained minimization
that leads to the formation of a resonance that describes the Kondo peak.

\section{Summary and Conclusions}

\label{Sec5}

We employed the X-boson technique to calculate the specific heat of the PAM in
two regimes (HF-K and HF-LMM), and we find agreement with recent theoretical
work\cite{Lacroix, Chinos,Portuges,PruschkeBJ,CostiM,BurdinGG} as well as
qualitative agreement with measurements\cite{Steglich1, Steglich2} of
Sommerfeld's coefficient $\gamma(T)=C_{v}(T)/T$. The low temperature entropy
shows a crossover from a singlet ground state in the HF-K regime to a lattice
of doublets in the HF-LMM region, and we discuss the origin of this behavior.
We conclude that our method gives fairly good results in the Kondo region not
very far from the intermediate valence region, as well as in this last region.

The specific heat in the HF-LMM region has a $C_{v}(T)\sim\beta T^{3}$
contribution, that can be attributed to Sommerfeld's second term in the
temperature expansion of $C_{V}$. It would be interesting to calculate the
electrical resistivity in the two regimes at low temperatures ($T<T_{0}$).

\begin{acknowledgments}
We would like to express our gratitude to Prof. Eduardo Miranda and to Prof.
Ben Hur Bernhard for several helpful discussions. The financial support of the
S{{\~{a}}}o Paulo State Research Foundation (FAPESP), the National Research
Council (CNPq) and the Latin American Center of Physics (CLAF) is gratefully acknowledged.
\end{acknowledgments}

\begin{figure}[ptb]
\begin{center}
\includegraphics[clip,width=0.40\textwidth,
height=0.30\textheight]{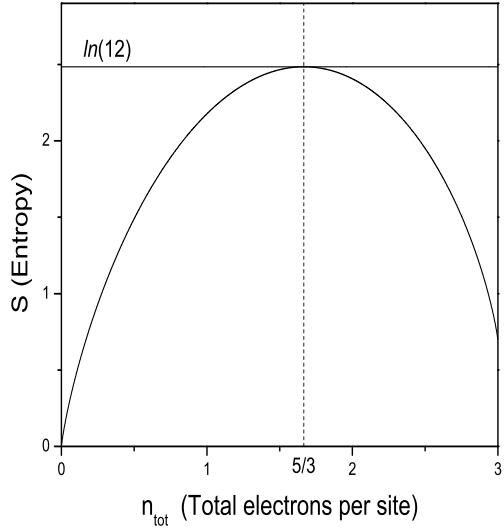}
\end{center}
\caption{Entropy per site of the PAM for $T\rightarrow\infty$ as a function of
the total number of electrons per site $y=n_{tot}$. The horizontal line
corresponds to $S_{T\rightarrow\infty}\left(  5/3\right)  =\ln\left(
12\right)  $.}%
\label{F3}%
\end{figure}

\appendix

\section{The entropy of the PAM for very high $T$}

\label{ApA}

When $T\rightarrow\infty$ one is tempted to consider a uniform occupation of
all the states, but we shall show that this is only true when the chemical
potential is kept constant. This assumption is not generally true for a system
with a fixed total number $n_{tot}=N_{tot}/$ $N_{S}$ of electrons per site
($N_{S}$ is the number of sites), and we shall calculate the entropy in this
limit as a function of $n_{tot}$.

When $T\rightarrow\infty$ we can set (we take $k_{B}=1$) $V/T=W/T=0$, and we
can write the thermodynamic potential per site%
\begin{align}
\frac{\Omega}{N_{S}}  &  =T\ \ln\left[  \left(  1+2\ \exp\left[  -\left(
E_{f}-\mu\right)  /T\right]  \right)  \right.  \ \nonumber\\
&  \times\left.  \left(  1+\exp\left[  +\mu/T\right]  \right)  ^{2}\right]  ,
\label{A1}%
\end{align}
and%
\begin{equation}
n_{tot}=\frac{2\ \exp\left[  -\left(  E_{f}-\mu\right)  /T\right]  }%
{1+2\ \exp\left[  -\left(  E_{f}-\mu\right)  /T\right]  }+\frac{2\ \exp\left[
+\mu/T\right]  \ }{1+\exp\left[  +\mu/T\right]  }. \label{A2}%
\end{equation}
For convenience we write%
\begin{align}
x  &  =\exp\left[  -\mu/T\right]  ,\nonumber\\
y  &  =n_{tot},\nonumber\\
e_{f}  &  =\exp\left[  E_{f}/T\right]  \label{A3}%
\end{align}
so that Eq. (\ref{A2}) can be written as%
\begin{equation}
y=\frac{2}{e_{f}\ x+2}+\frac{2}{x+1}. \label{A4}%
\end{equation}
As we are interested in the limit $T\rightarrow\infty$ we write here $e_{f}%
=1$, an solve $x$ as a function of $y$:%
\begin{equation}
x=\frac{\left(  4-3y+\sqrt{16+y^{2}}\right)  }{2y}. \label{A5}%
\end{equation}
The solution with $-\sqrt{16+y^{2}}$ gives negative values of $x$ and we have
to bound the total number of electrons $n_{tot}=y$ to the interval $[0,3]$,
i.e. only take%
\begin{equation}
0\leq y\leq3. \label{A6}%
\end{equation}

It is clear that except for $x=1$ (which corresponds to $n_{tot}=5/3$) we have
\ $\mu=\alpha\ T$, that goes to infinity together with $T$, and from Eq.
(\ref{A4}) or Eq. (\ref{A5}) follows that the value of $\alpha$ has the sign
of $\left(  n_{tot}-5/3\right)  $.

The thermodynamic potential and the entropy can now be calculated for
$T\rightarrow\infty$ as a function of $y=n_{tot}$. The expression for the
entropy is%
\begin{align}
&  S_{T\rightarrow\infty}\left(  y\right)  =y\ \ln\left[  \frac{4-3y+\sqrt
{16+y^{2}}}{2y}\right] \label{A7}\\
&  +\ln\left[  \frac{\left(  4-y+\sqrt{16+y^{2}}\right)  ^{2}\ \left(
4+y+\sqrt{16+y^{2}}\right)  }{\left(  4-3y+\sqrt{16+y^{2}}\right)  ^{3}%
}\right] \nonumber
\end{align}
and is plotted in figure \ref{F3}.\newline The entropy for \ $n_{tot}=1.6$ is
$S_{T\rightarrow\infty}\left(  1.6\right)  =0.998765\ \ln\left(  12\right)  $
(see figure \ref{fig3_CvL}) while for \ $n_{tot}=1.9$ is $S_{T\rightarrow
\infty}\left(  1.9\right)  =0.984597\ \ln\left(  12\right)  $ (see figure
\ref{fig9_CvL}). It follows from Eq. (\ref{A1}) that for $T\rightarrow\infty$
it is%
\begin{equation}
n_{f}=\frac{2}{2+x}\hspace{0.5cm}\mathrm{and\hspace{0.5cm}}n_{c}=\frac{2}%
{1+x}, \label{A8}%
\end{equation}
where as before \ $x=\exp\left[  -\mu/T\right]  $.

\end{document}